# Scholte waves generated by seafloor topography

*Yingcai Zheng, Xinding Fang, Jing Liu and Michael C. Fehler\*, Department of Earth, Atmospheric, and Planetary Sciences, Massachusetts Institute of Technology.*

**Summary**

Seafloor topography can excite strong interface waves called Scholte waves that are often dispersive and characterized by slow propagation but large amplitude. This type of wave can be used to invert for near seafloor shear wave velocity structure that is important information for multi-component P-S seismic imaging. Three different approaches are taken to understand excitation of Scholte waves and numerical aspects of modeling Scholte waves, including analytical Cagniard-de Hoop analysis, the boundary integral method and a staggered grid finite difference method. For simple media for which the Green's function can be easily computed, the boundary element method produces accurate results. The finite difference method shows strong numerical artifacts and stagnant artificial waves can be seen in the vicinity of topography at the fluid-solid interface even when using fine computational grids. However, the amplitude of these artificial waves decays away from the seafloor. It is sensible to place receivers away from the fluid-solid interface for seismic modeling. To investigate Scholte wave generation, one needs to correctly implement the fluid-solid boundary condition. It is also shown through numerical examples including using a seafloor profile from the recent SEG Advanced Modeling (SEAM) Project that even mild topographic features can generate Scholte waves and these waves can be used to constrain near seafloor S wave velocity by dispersion analysis of interface Scholte waves. The implication to the full waveform inversion is that, although low frequency data are crucial for convergence, seafloor topography may have a large effect on low frequency seismic waves.

**Introduction**

One of the main benefits of multicomponent ocean bottom seismic acquisition is to record P-to-S converted waves, which can be used to obtain better reservoir images for characterization. However, some new processing difficulties are also introduced. One outstanding issue is the slow but large-amplitude dispersive Scholte wave that propagates along the seafloor. The Scholte wave is like ground-roll but at the seafloor and it can possibly obscure and mask deep reflections from the reservoir. The second outstanding issue is estimating the near seafloor shear wave velocity. It has been shown that the shear velocity near the seafloor can be as low as ~10m/s (Jackson and Richardson, 2007). This will cause significant slow down for upgoing P-S conversions near the seafloor. In high-resolution depth imaging, it is very important to accurately estimate the shallow shear wave velocity structure and correct for the statics. Scholte waves have a benign side and they have been commonly used to infer shear wave velocity just below the seafloor by fitting their dispersion curves (Tango et al., 1986; van Vossen et al., 2002; Bohlen et al., 2004; Kugler et al., 2005; Kugler et al., 2007; Muyzert, 2007a; b). The generation of Scholte waves is similar to that of Rayleigh waves by a point source below a free surface and is strongly dependent upon the distance between the source and the interface. Strong Scholte waves can be generated when the source is close to the seafloor. For sources further from the seafloor, only low frequency and weak Scholte waves can be excited directly by the source. However, there is another mechanism to effectively excite Scholte waves, which is by seafloor topography. The seafloor is not necessarily flat and this feature has been captured by the SEG Advanced Modeling (SEAM) Project (Fehler, 2012). Accurate modeling of the wave propagation through/along a rough seafloor is important for better seismic imaging. In this abstract we first study not only properties of Scholte waves and their generation by topography but we also investigate numerical artifacts of using a staggered grid finite difference method compared to the more accurate boundary element method, which faithfully satisfies the fluid-solid boundary condition.

**Theory**

For a planar interface, the generation of Scholte wave has been well understood theoretically (de Hoop and van der Hijden, 1983; de Hoop and van der Hijden, 1984). We assume that the fluid velocity is $c_f$ and density $\rho_f$; the underlying solid has $c_P$, $c_S$ and $\rho_S$. Leaky Rayleigh waves and Scholte waves can exist along the interface and they satisfy

$$\frac{1}{4c_s^4}\frac{\rho_f \eta_p}{\rho_s \eta_f} + \left(p^2 - \frac{1}{2c_s^2}\right)^2 + p^2 \eta_p \eta_s = 0 \qquad (1)$$

where $p$ is the complex slowness and

$$\eta_{f,p,s} = \sqrt{c_{f,p,s}^{-2} - p^2} \qquad (2)$$

and

$$\mathrm{Re}(\eta_{f,p,s}) \geq 0. \qquad (3)$$

Equation (1) has eight Riemann sheets owing to three square roots. It has been shown that the Scholte wave always exists for a fluid-solid interface and its velocity

# Scholte wave generation by seafloor topography

$c_{sch}$ is less than all body wave velocities, i.e., $c_{sch} < \min(c_p, c_S, c_f)$.

If the seafloor has topography, one good way to solve the wave propagation problem is to use the boundary integral equation method (e.g., Zheng, 2010). For the direct boundary element modeling (BEM), we essentially solve the following coupled integral equations formulated in the frequency domain on the seafloor. In the solid, we have

$$\frac{1}{2}u_k(\mathbf{x}') = -\iint u_i(\mathbf{x})\Sigma_{ij}^k(\mathbf{x}|\mathbf{x}')n_j(\mathbf{x})dS(\mathbf{x}) \\ + \iint G_{ik}(\mathbf{x}|\mathbf{x}')t_i(\mathbf{x})dS(\mathbf{x}), \mathbf{x}' \in S \quad (4)$$

where $i, j, k = 1$ and 3 in 2D; $u_i$ and $t_i$ are displacement and traction unknowns on the boundary; $G_{ij}$ and $\Sigma_{ij}^k$ are elastodynamical Green and Green-traction tensors respectively (Sanchez-Sesma and Campillo, 1991); $n_i$ is the outward normal from the solid to the fluid at $\mathbf{x}$ on the boundary. In the fluid,

$$\frac{1}{2}p(\mathbf{x}') = p_0(\mathbf{x}') - \iint p(\mathbf{x})\frac{\partial G(\mathbf{x}|\mathbf{x}')}{\partial n(\mathbf{x})}dS(\mathbf{x}) \\ + \iint G(\mathbf{x}|\mathbf{x}')\frac{\partial p}{\partial n}(\mathbf{x})dS(\mathbf{x}), \mathbf{x}' \in S \quad (5)$$

in which $p$ and $\frac{\partial p}{\partial n}$ are the unknown pressure and its normal gradient on the boundary; $G$ is the Green's function in the fluid; $n$ is the outward normal from the fluid to the solid. The boundary conditions between the solid and the fluid are continuity of normal displacement and traction

$$\frac{\partial p}{\partial n} = \omega^2 \rho_f n_k u_k \\ p = -n_j t_j \quad (6)$$

and vanishing tangential traction

$$\mathbf{n} \times \mathbf{t} = 0 . \quad (7)$$

To solve equations (4-7), we first discretize the 2D seafloor profile into linear segments. We assume that on each segment, $\mathbf{u}$, $\mathbf{t}$, $p$ and $\partial p/\partial n$ are constant. In fact, if we have $M$ segments, we will have $3M$ unknown boundary values and this will result in a large linear algebra system. The matrix is usually dense. Solving the matrix is challenging. However, iterative solvers and fast matrix-vector multiplication algorithms can be used. For the examples in this abstract, we use LU decomposition due to the small size of the problems. Once we have solved for the boundary values $\mathbf{u}$, $\mathbf{t}$, $p$ and $\partial p/\partial n$, we can use the Kirchhoff integral to calculate the wavefield at any position within the model.

**Examples**

In what follows, we will present several examples through which we can see that even small topographic features can generate Scholte waves. We will also point out some outstanding numerical issues in finite difference modeling due to improper treatment of the fluid-solid boundary condition.

Example 1. Our first example is for a seafloor with a small slope (< 3 degree dip) (Figure 1a). We use water velocity 1.5km/s and density 1.0g/cm$^3$. For the solid beneath the water, the compressional and shear velocities are $c_P = 1.7$ km/s and $c_S = 0.8$ km/s, respectively and the density is $\rho_S = 2.0$ g/cm$^3$. For these medium parameters, the Scholte pole is at $(1.455, 0)$ which corresponds to a propagation velocity of $c_{sch} \sim 0.682$ km/s. The source is an explosion with a Ricker source time function of central frequency $f_0 = 5.0$ Hz, which corresponds to a wavelength of 300 meters in water. There is no free surface. We use a standard 2D staggered-grid finite difference (FD) code with 4th order accuracy in space and 2nd order accuracy in time to simulate the seismic wave propagation at the seafloor. The bandwidth in the BEM modeling is from 0 Hz to 15Hz. Salient seismic phases include head P wave, direct P wave and strong low frequency Scholte wave excited by the explosion source (Figure 1b & c). The lack of high frequency for the Scholte wave is due to the distance between the source and the seafloor. However, the slope changes at $x = \pm 2$ km also generate Scholte waves but at much higher frequencies compared to the source generated Scholte waves. Due to the direction of the incident wavefield, topography-generated Scholte waves propagating to the right have stronger amplitudes than to the left. In addition to the Scholte interface wave, we also observe small P waves generated by the topography. We have to keep in mind that the slope of the topography in this example is very small. For the staggered differencing scheme, there is no numerical stability issue if the shear modulus is set to be zero. Frequently, this is how wave propagation through/along a rough seafloor is modeled. In our FD modeling, the grid size is 5 meters. Regular FD meshing cannot be used to model the seafloor topography accurately and it has the well-known staircase effect (Figure 1b), which introduces numerical artifacts into the recorded signals. But those numerical artifacts decay away from the seafloor (Figure 2). In order to reduce the effect of numerical noise, receivers need to be placed at least a few grid points away from the seafloor.

# Scholte wave generation by seafloor topography

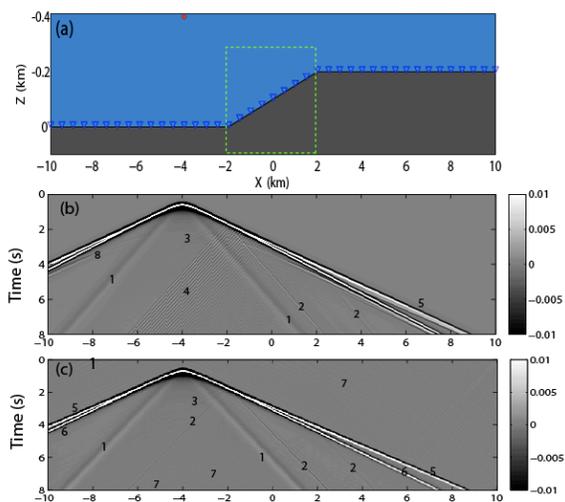

Figure 1. (a) A seafloor model which represents a gentle slope (<3 degrees); the red star is the source position and blue inverted triangles are geophones suited 10 meters above the seafloor; (b) pressure shot gather computed using fourth-order staggered-grid finite difference method; (c) shot gather computed using boundary element method. In (b) and (c), numbers within the plots designate different seismic phases. 1 source generated Scholte waves; 2 topography generated Scholte waves; 3 topography generated P waves; 4 finite difference method staircase numerical artifacts; 5 head P wave traveling beneath the seafloor; 6 direct P wave in water; 7 numerical noise in the BEM due to model truncation and wrap-around..

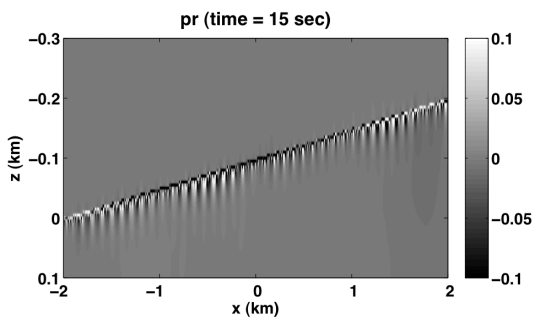

Figure 2. Snapshot of numerical artifacts from FD modeling at time of 15 s. This area corresponds to the area within the green dashed box in Figure 1a.

Example 2. The SEG Advanced Modeling (SEAM) (Fehler, 2012) project aims to build a realistic elastic deepwater model including seafloor topography and low shear velocities in unconsolidated marine sediments and to simulate seismic wave propagation through this model. The SEAM model has spatially variable velocity and density structures. We use one SEAM seafloor profile to study whether Scholte waves can be generated by seafloor topography (Figure 3). Below the water, we use a homogeneous elastic model, $c_P = 1.8$ km/s, $c_S = 0.8$ km/s and $\rho_S = 2.0$ g/cm$^3$. The source is located at $(x = 20\text{km}, z = 0)$ (Figure 3). The center frequency for the Ricker source wavelet is 2.0Hz. The receivers are placed 10 meters above the seafloor. Significant progress has been made in the full waveform inversion during last decade (Virieux and Operto, 2009). The general consensus is to have low frequency data to start with. However, low-frequency Scholte waves are more persistent do low frequency seismic data are more affected by Scholte waves generated at changes in seafloor topography than are high frequencies. Therefore, correct handling of the fluid-solid boundary condition is critical. For this SEAM profile, the topography is mild. The finite differnce code uses a 2.5m grid interval in the simulation and yet it produces almost stagnant noise having hyperbolic movout whose amplitude can be a few percent of the direct wave. This noise may mask out useful reflections from deeper structures.

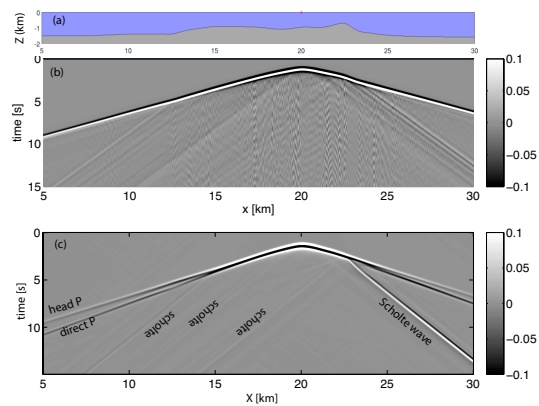

Figure 3. (a) SEAM topography profile; (b) shot gather from finite difference modeling normalized to the maximum amplitude; (c) shot gather by BEM normalized to the maximum amplitude.

Example 3. Scholte waves generated by a Gaussian hill (Figure 4a). The main frequency for the Ricker source wavelet is $f_0 = 3.0$ Hz. The wavelength is 500m. The topographic height of the Gaussian hill is 200m. The half width of the Gaussian hill is 200m side to side. $c_P = 1.7$ km/s, $c_S = 0.8$ km/s and $\rho_S = 2.0$ g/cm$^3$. Large-amplitude Scholte waves are excited by both incident P waves and by an earlier Scholte wave that was excited by the explosion source. It is possible to perform multi frequency dispersion analysis to determine the S-wave speed for marine sediments , particularly if there is a velocity gradient in the sediment. In this model, dispersion is not well developed because a majority of the model is homogeneous. We

# Scholte wave generation by seafloor topography

conducted some group velocity measurements. We used two station recordings to calculate the group velocity dispersion curve. We first select two recordings and isolate the interface Scholte wave. Next, we choose 10 central frequencies to perform Gaussian filtering to the two seismograms (Bensen *et al.*, 2007). For each frequency, we crosscorrelate the two filtered recordings and obtain the time lag, from which we calculate the group velocity between the two stations (Figure 5). The obtained group velocities are very close to the theoretical Scholte wave speed (~682m/s based on the Scholte pole position).

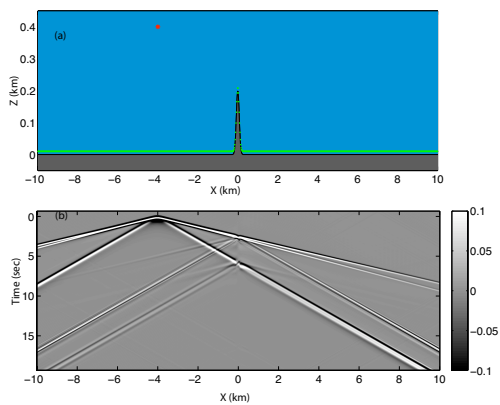

Figure 4. (a) A Gaussian hill model; the red star is the source position and green dots are receivers 10 meters above the seafloor; (b) common shot gather (pressure) computed using BEM and amplitudes are normalized with respect to the largest value for the whole gather.

Example 4. Random seafloor roughness model. We generate a random boundary having a Gaussian correlation function $\exp(-r^2/a^2)$. The horizontal correlation length is $a = 100$ m. The maximum peak-trough vertical distance is 200m (Figure 6a). The source is at x=7km and z=0km. Scholte wave excitation is strongly frequency dependent. For the same seafloor, we use two different sources, a 3Hz Ricker (Figure 6b) and a 5Hz Ricker (Figure 6c). Topography generated Scholte waves genered by the low frequency source last longer than those generated by the high frequency source (Figure 6).

## Conclusions

From both analytical and numerical analyses using BEM and FD modeling of Scholte waves, we see that Scholte waves are easily generated by a variety of topographic features. We also find that standard staggered grid finite difference schemes produce erroneous strong waves at locations where topography changes. These waves do not seem to propagate far but remain localized. Topography generated multi-frequency Scholte waves can be clearly identified and their dispersion characteristics can be used to estimate near seafloor shear wave velocity structures which are important for multicomponent seismic imaging.

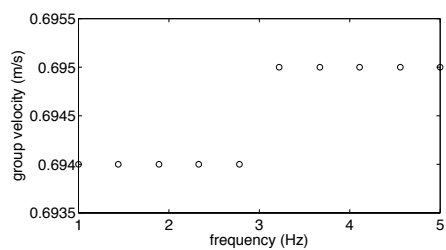

Figure 5. Group velocity between receiver $x = -2.49$ km and $x = 5.02$ km for the Scholte wave propagating to the right in Figure 4b.

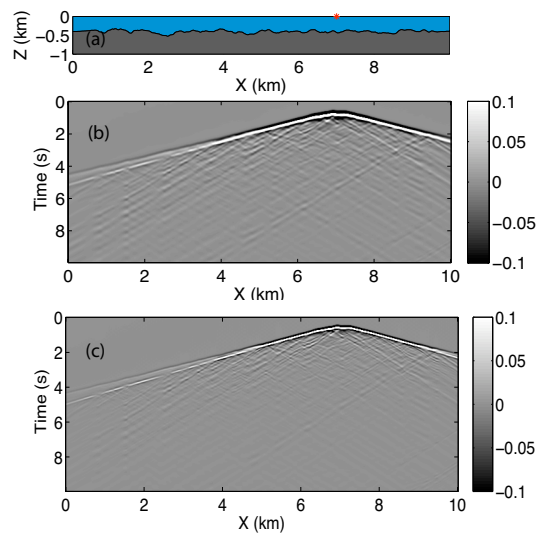

Figure 6. (a) A random Gaussian boundary; (b) shot gather for 3Hz Ricker source; (c) shot gather for 5Hz Ricker source. BEM is used to compute seismograms.

## Acknowledgments

This work is supported by ERL/MIT Consortium. We thank SEAM for allowing us to use their seafloor topography model.

References

Bensen, G. D., M. H. Ritzwoller, M. P. Barmin, A. L. Levshin, F. Lin, M. P. Moschetti, N. M. Shapiro, and Y. Yang (2007), Processing seismic ambient noise data to obtain reliable broad-band surface wave dispersion measurements, *Geophys. J. Int.*, **169**(3), 1239-1260.

Bohlen, T., S. Kugler, G. Klein, and F. Theilen (2004), 1.5D inversion of lateral variation of Scholte-wave dispersion, *Geophysics*, **69**(2), 330-344.

de Hoop, A. T., and J. H. M. T. van der Hijden (1983), Generation of acoustic waves by an impulsive line source in a fluid/solid configuration with a plane boundary, *Journal of the Acoustical Society of America*, **74**(1), 333-342.

de Hoop, A. T., and J. H. M. T. van der Hijden (1984), Generation of acoustic waves by an impulsive point source in a fluid/solid configuration with a plane boundary, *J. Acoust. Soc. Am.*, **75**(6), 1709-1715.

Fehler, M. (2012), SEAM update: Elastic simulations, *The Leading Edge*, **31**(1), 24-25.

Jackson, D., and M. Richardson (2007), *High-Frequency Seafloor Acoustics,* Springer, New Youk (Chapter 5).

Kugler, S., T. Bohlen, S. Bussat, and G. Klein (2005), Variability of Scholte-wave dispersion in shallow-water marine sediments, *J Environ Eng Geoph*, **10**(2), 203-218.

Kugler, S., T. Bohlen, T. Forbriger, S. Bussat, and G. Klein (2007), Scholte-wave tomography for shallow-water marine sediments, *Geophys. J. Int.*, **168**(2), 551-570.

Muyzert, E. (2007a), Seabed property estimation from ambient-noise recordings: Part 2 - Scholte-wave spectral-ratio inversion, *Geophysics*, **72**(4), U47-U53.

Muyzert, E. (2007b), Seabed property estimation from ambient-noise recordings: Part I - Compliance and Scholte wave phase-velocity measurements, *Geophysics*, **72**(2), U21-U26.

Sanchez-Sesma, F. J., and M. Campillo (1991), Diffraction of P, SV, and Rayleigh waves by topographic features: a boundary integral formulation, *Bull. Seismol. Soc. Am.*, **81**(6), 2234-2253.

Tango, G. J., W. J. Cafarelli, and H. Schmidt (1986), Sediment Shear Profiling Using Scholte Waves - Numerical-Experimental Study in Breton Sound, Offshore Louisiana, *Geophysics*, **51**(6), 1324-1324.

van Vossen, R., J. O. A. Robertsson, and C. H. Chapman (2002), Finite-difference modeling of wave propagation in a fluid-solid configuration, *Geophysics*, **67**(2), 618-624.

Virieux, J., and S. Operto (2009), An overview of full-waveform inversion in exploration geophysics, *Geophysics*, **74**(6), WCC1-WCC26.

Zheng, Y. (2010), Retrieving the exact Green's function by wavefield crosscorrelation, *J. Acoust. Soc. Am.*, **127**(3), EL93-EL98.